\documentclass{article}
\usepackage[french]{babel}
\usepackage{setspace} 
\usepackage[T1]{fontenc}
\usepackage{amsmath}
\usepackage{amsfonts}
\usepackage{amssymb}
\newtheorem{defi}{Définition}
\newtheorem{teo}{Théorème}

\begin{document}

\title{\LARGE{Sur l'universalité des classes\\ de calcul des nombres réels}}
\author{Héctor Zenil\footnote{hector.zenil-chavez@malix.univ-paris1.fr}\\
IHPST, Universit\'e de Paris I (Pantheon-Sorbonne)}
\maketitle

\abstract{Les modèles de calcul qui opèrent sur les ensembles de nombres réels, et qui sont capables de calculer sur une classe strictement supérieure à celle des fonctions récursives, introduisent invariablement un élément non fini, soit comme information encodée dans une entrée, soit comme information contenue dans le modèle. Dans cet article, on montrera que la propriété d'universalité des modèles de calcul est liée aux degrés de Turing des maximaux atteints par les éléments en jeu, en particulier le domaine d'opération et les fonctions permises. Pour le cas le plus général des modèles qui opèrent sur l'ensemble complet des nombres réels, cette propriété d'universalité au sein du concept de calcul, se limite seulement aux coupes arbitraires de la hiérarchie arithmétique. Cette propriété d'universalité permet d'établir un domaine non borné d'opération et des fonctions de puissance arbitraire en parcourant tous les degrés de calcul. Elle accepte donc des modèles ouverts, sous leurs propres opérations. Une définition d'universalité relative, d'universalité interne et d'universalité
externe est proposée.
\\

Mots clés: Universalité, machines de Turing, fonctions récursives, calculabilité, nombres réels.}

\section{Contexte}

Certains chercheurs ont proposé des modèles de calcul avec la capacité
d'opérer sur des nombres réels, ces modèles peuvent être trouvés sous des appellations comme l'"ordinateur différentiel" ou l'"informatique analogique".
Même le concept de ce qui est aléatoire dans la théorie algorithmique de l'information est fondé sur l'idée de l'existence d'une partie non dénombrable ou la totalité des nombres réels, c'est-à-dire le continu. \\

L'une des propriétés les plus importantes de la classe des machines de Turing \cite{turing} est celle de l'universalité. En effet, cette notion se trouve au coeur de la théorie du calcul et de la théorie de la complexité qui donne lieu à la conception de l'ordinateur à usages multiples. \\

Au sein des modèles traditionnels (dits modèle de Turing et équivalents), l'universalité est possible grâce à la codification de la table des actions de n'importe quelle machine en une chaîne finie de symboles. Donc il est possible de construire une machine de Turing $U$ qui reçoive comme entrée dans son ruban la table des actions de la machine $M$. Cette machine $U$ va ainsi recevoir l'entrée $x$ de $M$, qui est codifiée pour calculer avec $U$ ce que la machine $M$ aurait calculé. \\

Donc, l'existence d'une machine universelle de Turing est fondée sur la possibilité d'énumérer toutes les machines de Turing. Mais même si l'énumération semble être une condition nécessaire, elle ne suffit pas, car des modèles d'automates moins puissants sont évidemment dénombrables, néanmoins ils n'admettent pas d'universalité \footnote{En effet, il n'existe pas d'automate fini universel $U$. Si on proposait un tel automate $U$ on aurait un dispositif dans la classe $L$ des automates finis capable de simuler n'importe quel autre automate $A$ de la même classe $L$. Alors qu'il est facile de construire un automate fini $A$$\prime$ qui accepte un langage $L_i$$\in$$L$$\prime$ en utilisant le lemme de "pumping"\cite{kozen} qui ne soit pas accepté par $U$ : ce qui devient une contradiction.}. \\

D'autre part, tous les modèles d'hyper-calcul semblent être de deux types \cite{cotogno} : Ceux qui utilisent une sorte d'"oracle" (analysé par Turing même) et ceux qui calculent un nombre infini d'opérations en temps fini. On peut ajouter que le pouvoir d'un modèle de calcul en général se base sur deux entités : a) l'information que l'automate est capable de contenir, soit codifiée $\mathit{a}$ $\mathit{priori}$ dans sa configuration, soit comme entrée.  Et b) l'ensemble des fonctions avec lesquelles il est capable d'opérer. En autres termes, ce que l'automate peut accepter, et ce que l'automate réalise en utilisant ses opérations. Les deux ensembles, dits de domaine $D$ et de fonctions $F$, déterminent la puissance du modèle de calcul. Dans le cas de machines de Turing, il est clair que $D$ est l'ensemble des langages récursivement énumérables, tandis que $F$ est l'ensemble des fonctions récursives. \\

Dans le cas des modèles d'"hyper-calcul" \footnote{La plupart des chercheurs\cite{davis2} pensent que la réalisation de ces machines n'est pas possible. Cependant, au-delà de cette discussion pertinente on s'est aperçu que ces modèles ont été importants pour mieux comprendre la théorie de la calculabilité et même le modèle des machines de Turing ou la théorie des fonctions récursives.}, ceux-ci semblent toujours avoir besoin qu'un élément non fini (complet), soit dans leur domaine, soit un
élément impliqué pendant leurs opérations (où la thèse de Church-Turing entre en jeu).\\

\section{Universalité relative et perte d'universalité}

Il faut définir d'abord les concepts d'automate, de classe et d'universalité en termes les plus généraux possibles pour que les résultats soient applicables à tous les modèles de calcul:

Si $A$ est un automate universel sur une classe $K$ d'automates avec des entrées en $X$ où $X$ est l'ensemble de toutes les chaînes de symboles d'un alphabet donné, $A$ peut se définir comme une fonction $A$:$X$$\rightarrow$$X$ quand les sorties appartiennent toujours au même ensemble $X$ des entrées. Dans ce cas, on dit que $A$ est fermé. Nous savons que $A$ peut être une fonction totale ou partielle. Si $A$ est une machine de Turing, $A$ est en général partielle étant donné le problème de l'arrêt, c'est-à-dire les entrées n'y sont pas nécessairement  associées à une sortie. Donc si $A$ est toujours total, $A$ est strictement moins puissant qu'une machine de Turing. On considéra plus tard le cas dans lequel $A$:$X$$\rightarrow$$Y$ tel que $X$$\subset$$Y$ comme dans certains cas non fermés des modèles d'"hyper-calcul" proposés.\\

Par ailleurs, une classe de calcul est définie par les classes de complexité de calcul des degrés de Turing\footnote{Le degré de Turing\cite{cooper} d'une classe $A$ est: $deg_T$$(A)$$=$\{$X$$|$$X$$\equiv$$_T$$A$\} où $A$$\equiv$$_T$$B$ ssi $A$$\leq_T$$B$ et $B$$\leq_T$$A$. En d'autres termes, le degré de Turing d'une classe $A$ des nombres naturels $\mathbb{N}$, c'est la classe d'équivalence de toutes les sous-classes de $\mathbb{N}$ sous la réductibilité de Turing. La réductibilité de Turing d'un problème $A$ à un problème $B$ est une réduction qui résout $A$ en supposant $B$. C'est-à-dire, une réduction de Turing est une fonction calculable par une machine de Turing avec un oracle pour $B$.} définis en termes d'oracles\footnote{Sous leur forme standard\cite{cooper}, ces machines possèdent un ruban spécial, qui est le ruban d'oracle, ainsi que trois états particuliers, $q_?$, $q_y$ et $q_n$. Le ruban d'oracle est un ruban d'écriture. Pour utiliser l'oracle, la machine écrit un mot sur ce ruban, puis va dans l'état $q_?$. Selon le mot, l'oracle décide si l'état suivant sera $q_y$ ou $q_n$.}.

Du travail de Turing\cite{turing} on a déduit que les langages peuvent être codés dans les expansions des nombres réels. Étant donné qu'on sait que pour toutes les chaînes d'un alphabet fini il existe un bon ordre. Donc il est toujours possible de codifier de façon unique un langage quelconque dans le développement d'un nombre réel $r$. Donc les propriétés du langage à codifier sont transférées au nombre qui le codifie, parce que c'est seulement une traduction. Un nombre rationnel sera donc capable de codifier un langage "rationnel", de la même façon qu'un nombre Turing-calculable ne sera pas capable de codifier un langage non Turing-calculable.\\

D'ailleurs il est clair que par le théorème de Post\cite{post}\footnote{$\forall$ $n$$>$$0$
\begin{enumerate}
\item 0$^{(n+1)}$ est $\Sigma_{n+1}$-complet, c'est-à-dire que le n+1-ième degré de Turing après , est $\Sigma_{n+1}$-complet (complet signifie dans ce cas qu'il est un représentant des problèmes les plus difficiles en complexité, comme dans le sens de NP-complet comme on verra plus tard).
\item $A$ est $\Sigma_{n+1}$ $\Leftrightarrow$ A est récursivement énumérable avec l'oracle 0$^{(n)}$
\item $A$ est $\Delta_{n+1}$ $\Leftrightarrow$ A $\leq_T$ 0$^{(n)}$
\end{enumerate}
} ces classes de calcul parcourent la hiérarchie arithmétique\footnote{$\Pi_n$ :  $\forall$$a_1$$\ldots$$\exists$$a_2$$\ldots$$\phi$($a_1$$a_2$$\ldots$$a_n$) 
et  $\Sigma_n$ :  $\exists$$a_1$$\ldots$$\forall$$a_2$$\ldots$$\phi$($a_1$$a_2$$\ldots$$a_n$) 
où $\phi$ ne contient pas de quantificateurs.\\
$\Delta_n$=$\Sigma_n$ $\cap$ $\Pi_n$.} pour sortir au niveau de l'hyper-hiérarchie arithmétique, de telle sorte qu'ils comprennent tous les types de complexité de calcul définissables par les nombres réels. On appelera sous-automates ou seulement automates, les automates au-dessous du modèle de Turing dans la hiérarchie d'automates, et on appellera hyper-automates les automates capables de résoudre le problème de l'arrêt des machines de Turing.\\

Maintenant nous sommes en position de continuer et de donner la définition précise d'universalité par rapport à une classe de calcul comme suit :

\begin{defi} \cite{burgin} Un automate $U$ est universel pour une classe $K$ si, étant donnée une $d(A)$ comme description d'un automate $A$ en $K$ et une entrée $x$ pour lui, $U$ est capable de se comporter pas à pas comme $A$ et donner le même résultat qu'aurait donné $A$ pour l'entrée $x$.
\end{defi} 

Il est donc clair que $U$ est capable de simuler n'importe quel automate de $K$ car $A$ est un automate quelconque.

\begin{defi} \label{a} Si par ailleurs $U$ est universel pour $K$ et appartient à $K$ alors $U$ est appelé un automate universel de $K$. Étant donné que tout automate universel U$\prime$ en $K$ est capable de simuler n'importe quel autre automate en $K$ comme le fait $U$, alors $U$ et U$\prime$ calculent le même ensemble de fonctions. On dira alors que la classe $K$ d'automates est intrinsèquement universelle car on a une classe d'automates en $K$ capables de calculer l'ensemble complet d'automates de $K$ -l'ensemble de fonctions récursives en $K$-.
\end{defi}

L'exemple le plus commun est celui des machines de Turing. Nous savons qu'il existe une machine de Turing universelle - en fait un ensemble de machines universelles - pour la classe de toutes les machines de Turing qui est aussi une machine de Turing\cite{cooper}.\\

Pour construire une machine de Turing universelle on a besoin de codifier toutes les machines de Turing possibles. Parce qu'elles sont dénombrables il est possible de concevoir une telle machine universelle.\\

À l'exception du ruban d'une machine de Turing, la description d'une telle machine est évidemment finie. Quant au ruban, même quand ses définitions semblent faire appel au concept d'infini potentiel il est aussi possible de le définir comme un ruban de telle sorte que même quand il est capable de sauver une quantité infinie d'information, la quantité d'information qui est réellement sauvée à chaque instant est toujours finie. Donc, la définition complète d'une machine de Turing pour chaque instant est finie. Et la définition complète d'une machine de Turing aussi car le ruban n'est jamais rempli $\mathit{a}$ $\mathit{priori}$ avant de commencer. Cependant, ce qui fait l'existence d'une machine universelle de Turing, c'est évidemment le ruban auquel elle a accès. Ce ruban est un espace abstrait non limité, ce qui permet de sauver et d'opérer sur une quantité potentiellement infinie d'information même dans les cas où il entre en cycles infinis : les  cas pour lequels le problème de l'arrêt apparaît.\\

De plus, une description $d(T)$ doit être une suite discrète de symboles puisqu'un autre type de chaîne ne serait pas accepté par $U$ car $U$ est aussi une machine de Turing décrite en termes finis.\\

Cependant, même quand $d(A)$ pour un automate fini $A$ est dénombrable et donc arithmétisable, elle ne suffit pas pour accueillir un automate fini pour la classe d'automates finis $L$. Bien que $U$, la machine universelle de Turing pour la classe de machines de Turing est en fait l'automate universel pour la classe des automates finis, il ne l'est pas dans le sens de Turing dans la définition \ref{a}, car $U$ n'appartient pas à $L$. Donc, il semble que l'arithmétisation des automates d'une classe $L$ soit une propriété nécessaire mais pas suffisante en général pour réaliser l'universalité intrinsèque de $L$. Pour résumer, les classes où on trouve des automates universels qui appartientiennent à la classe qu'ils simulent sont intrinsèquement universels.

\subsection{Complexité d'un automate}

En utilisant la liaison évidente entre les ensembles de nombres et les automates qui les calculent\cite{minsky,sieg}, on peut utiliser une mesure de complexité pour les automates, fondée de manière naturelle sur la complexité des nombres réels:

\begin{defi} Soit $A$ un automate d'une classe $L$ où $L$ permet les opérations $f_1$,$\ldots$,$f_n$$\in$$F$ avec $n$$\in$$\mathbb{N}$. Alors, la complexité de $A$, dénotée par $C(A)$ est le degré des éléments maximaux de Turing de l'ensemble des nombres que $A$ peut calculer.\end{defi}

L'ensemble des nombres que $A$ peut calculer est l'ensemble des nombres réels qui encodent les langages acceptés par $A$, qui sont encodés dans $d(A)$ en tant que description de $A$ et encodés par l'ensemble de fonctions $F$$\in$$A$.\\

Par exemple, $C(T)$=$C(q)$=$O\prime$ par définition, si $T$ est une machine de Turing, automate fini ou automate de pile. Car $q$ ne peut être qu'un nombre réel calculable et en utilisant le théorème de Post $C(q)$=$\Sigma_1$, $C(T)$ se trouve au fond de la hiérarchie arithmétique. \\

Cependant, si $C(A)$ est un hyper-automate $C(A)$=$C(r)$ pour un nombre réel $r$ quelconque même s'il n'est pas calculable. Donc $C(A)$=$C(r)$=$O^n$ étant donné que $A$ est capable de calculer n'importe quel nombre $r$. Il faut éclaircir d'ailleurs le fait qu'étant donné que les degrés de Turing ne forment pas un ordre linéaire, $r$ peut concerner l'ensemble des éléments maximaux des degrés de Turing des nombres réels en jeu en $A$, c'est-à-dire l'ensemble des nombres réels dotés d'une complexité majeure, en comparaison avec ceux qui peuvent être comparés avec eux. Ce qui ne change pas du tout la définition donnée de la complexité de $r$.\\

\begin{defi} Une classe $F$$>$$G$, si le degré de Turing de $F$ est majeur par rapport au degré de Turing de $G$, c'est-à-dire $deg_T(F)$$>$$deg_T(G)$. \end{defi}

En d'autres termes $F$ contient au moins un langage $f$ tel qu'il existe un automate $A$ de classe $F$ pour que $L$($A_F$)$=$$f$ tandis qu'il n'existe pas d'automate $B$ dans $G$ tel que $L$($B_G$)$=$$f$. C'est-à-dire que $A$ est un automate au pouvoir de calcul strictement majeur à $B$, et donc $F$ est une classe de calcul strictement majeure.

\section{Le calcul des nombres réels}

\begin{defi} Le langage généré par un automate $A$ sera décrit par $L(A)$ où $L(A)$$\in$$L$, quelle que soit la classe $L$ dont $A$ fait partie.
\end{defi} 

\begin{defi} Une classe $F$ est fermée sous un automate $A$ si le langage $L(A)$$\in$$F$, c'est-à-dire si le langage $L(A)$ produit par $A$ provisionné des opérations $f_i$,$\ldots$,$f_n$ avec $n$$\in$$\mathbb{N}$, fait partie de $F$.\end{defi}

Il est bien connu\cite{burgin} que les fonctions récursives préservent la fermeture des classes de calcul et qu'il existe des opérations topologiques ou à la limite qui ne le font pas.

\begin{teo} Soit $K$ et $K\prime$ deux classes de calcul tel que $K$$<$$K\prime$ alors $K$ et $K\prime$ sont des classes fermées sous les opérations des automates de chaque classe . C'est-à-dire, il n'existe pas d'automate $A$ en $K$ capable de calculer au niveau de $K\prime$, autrement $K$ s'effondrerait sur $K\prime$.
\end{teo}

\noindent Démonstration : C'est direct étant donné que les classes de calcul ici considérées sont fondées sur la hiérarchie arithmétique et les degrés de Turing.

\begin{teo} Soit $A$ un automate d'une classe fermée $K$ dans lequel on permet des éléments codifiés par nombres réels ou des nombres réels. Alors $A$ ne peut pas se décomposer en parties ni finies ni énumérables et $K$ ne permet pas d'automate universel dans $K$ lui-même.
\end{teo}

\noindent Démonstration (par contradiction) : Soit $N$ un automate universel dans la classe précédente $K$ et $C(N)$=$0^n$. Soit la classe $K$ une classe fermée qui permet des calculs de nombres réels arbitraires. Donc il est facile d'y construire $N\prime$ tel que $C(N\prime)$=$0^{n\prime}$ où $0^n$$<$$0^{n\prime}$, en choisissant les nombres réels appropriés selon ces degrés de Turing. Donc, $N$ ne peut pas être universel en $K$, car il n'accepte pas l'automate $N\prime$ en $K$.\\

En d'autres termes, parce que les nombres réels ne sont pas bornés en complexité, les modèles de calcul utilisant des nombres réels ne permettent pas l'existence d'un automate universel.

\section{Modèles de calcul non-fermés}

Prenons maintenant le cas le plus général des classes au-delà de la classe des machines de Turing:

\begin{teo} (le résultat principal): Si $A$ est un automate d'une classe $L$ fermée sous $A$ avec $L$$>$$T$ où $T$ est la classe des machines de Turing (la classe des ensembles récursivement énumerable) alors $L$ n'admet pas d'automate $U$ intrinsèquement universel\footnote{Il s'agit de la perte de la propriété d'universalité ; ce qui a des conséquences au niveau des fondements de la théorie du calcul, ainsi qu'àu niveau philosophique.}.
\end{teo} 

\begin{defi} Si $A$ est un automate de classe $M\prime$$>$$M$ et $A$ est un automate universel pour $M$, on dira que $M$ est  extrinsèquement universel et son automate extrinsèquement universel sera $A$. 
\end{defi}

\begin{teo} (deuxième résultat principal) Soit $M$ une classe quelconque, alors il existe un automate $A$$\in$$M$$+1$ tel que $A$ est universel pour $M$ où $M$$+$$1$ est la classe immédiate supérieure de $M$ selon les degrés de Turing.
\end{teo}

\noindent Démonstration : C'est tout de suite vrai pour le cas des automates de pouvoir inférieur ou égal à la classe des machines de Turing. En d'autres termes, il existe un automate $A$ de classe supérieure $M\prime$ aux machines de Turing qui simule n'importe quelle machine de Turing $T$ et $M\prime$$=$$M$$+$$1$, c'est-à-dire l'immédiate supérieure. Car $M\prime$ est la machine de Turing de premier oracle, elle résoud par définition le problème de l'arrêt des machines de Turing qui définit l'ensemble récursivement énumerable. Le même argument peut être utilisé aussi pour le reste des classes fondées sur les degrés de Turing, car chaque nouveau niveau résoud le problème de l'arrêt de la classe immédiat supérieure. Donc, la machine universelle cherchée est celle équivalente en pouvoir à la machine universelle de Turing d'oracle suivant.\\

L'universalité extrinsèque est une universalité qui n'est pas celle dans le sens de Turing. On a démontré qu'il est toujours possible d'atteindre un type d'universalité de telle sorte que pour n'importe quelle complexité d'un automate dans un certain modèle de calcul il y aura toujours un automate dans un modèle plus puissant capable de le simuler. \\

\section{$\Sigma_n$/$\Pi_n$$-$Universalité}

Il est possible de réaliser l'universalité au niveau de l'hyper-calcul en restreignant l'ensemble des nombres réels sur lequel le modèle opère, de façon à obtenir un type d'universalité par couches, défini  en termes de la hiérarchie arithmétique.

\begin{teo} Un modèle de calcul fermé sous l'union de son domaine d'opération et de ses fonctions, défini sur une classe $K$ tel que $C(K)$=$\Sigma_n$/$\Pi_n$, permet un automate intrinsèquement $\Sigma_n$/$\Pi_n$-universel en $K$.
\end{teo}
\noindent Démonstration : On construit un automate $U$:$X$$\rightarrow$$X$ comme suit : étant donné que $C(K)$=$\Sigma_n$/$\Pi_n$ on construit $U$ un automate tel que $r$$\in$$L(A)$ et $C(r)$=$\Sigma_n$/$\Pi_n$. On sait alors que $A$, en utilisant $r$, est capable de générer n'importe quel autre $r^\prime$ si $C(r)$$\geq$$C(r^\prime)$ et comme le modèle de calcul est fermé sous ses opérations il n'existe aucune façon d'atteindre des nombres réels d'une complexité majeure en passant par ses fonctions.\\

Il se trouve alors que pour chaque niveau de la hiérarchie arithmétique on a un modèle de calcul universel. En fait on a une quantité infinie dénombrable de modèles de calcul $\Sigma_n^m$/$\Pi_n^m$-universels.
\\

D'autre part, si le modèle n'est pas fermé soit sous ses fonctions soit par son domaine d'opération il n'admettra pas d'universalité intrinsèque.

\section{Conclusion}

Ce qu'on a démontré dans cet article, c'est que si les domaines des modèles de calcul sont de deux types, ceux qui sont fondés sur des éléments discrets et finis, et ceux qui utilisent le continu complet, la propriété d'universalité intrinsèque ne se trouve qu'au sein de la classe des machines de Turing; c'est ce qui la rend encore plus importante si on la compare aux autres modèles qui ne trouvent leur puissance que par l'usage de l'infini. Comme on a montré que la seule manière d'admettre l'universalité au niveau d'une classe de calcul $K$ se divise donc en trois possibilités:

\begin{enumerate}
\item Ou bien $K$ est la classe définie par les machines de Turing, où les opérations sont récursives et la classe est fermée.
\item  Ou bien si la classe de calcul $K$ est majeure par rapport à celle des machines de Turing, on a alors besoin de faire des coupes au niveau de la hiérarchie et de l'hyper-hiérarchie arithmétique pour obtenir des automates universels à chaque niveau.
\item  Ou bien on accepte des modèles non fermés sous leurs opérations, c'est-à-dire des opérations qui effondraient une partie ou la totalité de la hiérarchie arithmétique et des degrés de Turing.
\end{enumerate}

\bibliographystyle{latex8}

\end{document}